# Social Media, News and Political Information during the US Election: Was Polarizing Content Concentrated in Swing States?

Philip N. Howard
Oxford Internet Institute
Oxford University
philip.howard@oii.ox.ac.uk
@pnhoward

Bence Kollanyi
Oxford Internet Institute
Oxford University
bence.kollanyi@oii.ox.ac.uk
@bencekollanyi

Samantha Bradshaw
Oxford Internet Institute
Oxford University
samantha.bradshaw@oii.ox.ac.uk
@sbradshaww

Lisa-Maria Neudert
Oxford Internet Institute
Oxford University
lisa-maria.neudert@oii.ox.ac.uk
@lmneudert

**ABSTRACT**
*US voters shared large volumes of polarizing political news and information in the form of links to content from Russian, WikiLeaks and junk news sources. Was this low quality political information distributed evenly around the country, or concentrated in swing states and particular parts of the country? In this data memo we apply a tested dictionary of sources about political news and information being shared over Twitter over a ten day period around the 2016 Presidential Election. Using self-reported location information, we place a third of users by state and create a simple index for the distribution of polarizing content around the country. We find that (1) nationally, Twitter users got more misinformation, polarizing and conspiratorial content than professionally produced news. (2) Users in some states, however, shared more polarizing political news and information than users in other states. (3) Average levels of misinformation were higher in swing states than in uncontested states, even when weighted for the relative size of the user population in each state. We conclude with some observations about the impact of strategically disseminated polarizing information on public life.*

## COMPUTATIONAL PROPAGANDA AND THE 2016 US ELECTION

Social media plays an important role in the circulation of ideas about public policy and politics. Political actors and governments worldwide are deploying both people and algorithms to shape public life.[1,2] Bots are pieces of software intended to perform simple, repetitive, and robotic tasks. They can perform legitimate tasks on social media like delivering news and information—real news as well as junk—or undertake malicious activities like spamming, harassment and hate speech. Whatever their uses, bots on social media platforms are able to rapidly deploy messages, replicate themselves, and pass as human users. They are also a pernicious means of spreading junk news over social networks of family and friends.

Computational propaganda flourished during the 2016 US Presidential Election. There were numerous examples of misinformation distributed online with the intention of misleading voters or simply earning a profit. Multiple media reports have investigated how "fake news" may have propelled Donald J. Trump to victory.[3–5] What kinds of political news and information were social media users in the United States sharing in advance of voting day? How much of it was extremist, sensationalist, conspiratorial, masked commentary, fake, or some other form of junk news? Was this misleading information concentrated in the battleground states where the margins of victory for candidates had big consequences for electoral outcomes?

## SOCIAL MEDIA AND JUNK NEWS

Junk news, widely distributed over social media platforms, can in many cases be considered to be a form of computational propaganda. Social media platforms have served significant volumes of fake, sensational, and other forms of junk news at sensitive moments in public life, though most platforms reveal little about how much of this content there is or what its impact on users may be. The World Economic Forum recently identified the rapid spread of misinformation online as among the top 10 perils to society.[6] Prior research has found that social media favors sensationalist content, regardless of whether the content has been fact checked or is from a reliable source.[7] When junk news is backed by automation, either through dissemination algorithms that the platform operators cannot fully explain or through political bots that promote content in a preprogrammed way, political actors have a powerful set of tools for computational propaganda.[8] Both state and non-state political actors deliberately manipulate and amplify non-factual information online.

Fake news websites deliberately publish misleading, deceptive or incorrect information purporting to be real news for political, economic or cultural gain.[9] These sites often rely on social media to attract web traffic and drive engagement. Both fake news websites and political bots are crucial tools in digital propaganda attacks—they aim to influence conversations, demobilize opposition and generate false support.

## SAMPLING AND METHOD

Our analysis is based on a dataset of 22,117,221 tweets collected between November 1-11, 2016, that contained hashtags related to politics and the election in the US. Our previous analyses have been based on samples of political conversation, over Twitter that used hashtags that were relevant to the US election as a whole.

In this analysis, we selected users who provided some evidence of physical location across the United States in their profiles. Within our initial sample, approximately 7,083,691 tweets, 32 percent of the total traffic, were coming from users who volunteered enough location information in their profile to determine which state they were in. Many users have not provided enough details to allow us to identify which state they were tweeting from, added a location with a spelling mistake, or just entered a made-up location or a joke. After cleaning the data, we were able to successfully label 49.1 percent of the tweets from the users who provided any form of location information within their profile. The rest of the tweets were coming from profiles with a location field that refers to a location outside of the US or contains a location that we were not able to identify. Full access to user location data would provide better resolution on the distribution of users across states, but this response rate is sufficient to help us begin to understand the relationship between where users are and what kinds of political content they are being served by Twitter.

These tweets and associated data were collected from Twitter's public Streaming API at the time of the election, not retroactively with the Search API. The platform's precise sampling method is not known, but the company itself reports that the data available through the Streaming API is at most one percent of the overall global public communication on Twitter at any given time.[10] Tweets were selected based on a list of hashtags associated with the US election, and tweets were collected from the API that (1) contained at least one of the relevant hashtags; (2) contained the hashtag in the text of a link, such as a news article, shared in that tweet; (3) were a retweet of a message that contained the hashtag in the original message; or (4) quoted tweets in which the hashtag was included but in which the original text was not included and Twitter used a URL to refer to the original tweet.

To evaluate different sources being shared over social media, we determined the source of each of the URLs in the dataset. Overall, 1,275,430 of the 7,083,691 tweets from users that had provided location information also contained a URL. Many sources that were easy to identify were auto-coded, but other kinds of content obscured by link shorteners or re-shared less were also catalogued. Effectively this typology is built on successful cataloguing of 81% of the 1,275,430 links, with the remainder being single URLs shared only a few times or otherwise inaccessible. This typology that has emerged over our study of elections in five democracies:

- Professional News Outlets.
  o   Major News Brands. This is political news and information by major outlets that display the qualities of professional journalism, with fact-checking and credible standards of production. They provide clear information about real authors, editors, publishers and owners, and the content is clearly produced by an organization with a reputation for professional journalism. This content comes from significant, branded news organizations, including any locally affiliated broadcasters.
  o   Minor News Brands. As above, but this content comes from small news organizations or startups that display evidence of organization, resources, and professionalized output that distinguishes between fact-checked news and commentary. Professional publishers that cover other domains of social life, such as sports, fashion, or technology, but occasionally produce high quality political news.

- Professional Political Content
  o   Government. These links are to the websites of branches of government or public agencies.
  o   Experts. This content takes the form of white papers, policy papers, or scholarship from researchers based at universities, think tanks or other research organizations.
  o   Political Party or Candidate. These links are to official content produced by a political party or candidate campaign.

- Polarizing and Conspiracy Content
  o   Junk News. This content includes various forms of propaganda and ideologically extreme, hyper-partisan, or conspiratorial political news and information. Much of this content is deliberately produced false reporting. It seeks to persuade readers about the moral virtues or failings of organizations, causes or people and presents commentary as a news product. This content is produced by organizations that do not employ professional journalists, and the content uses attention grabbing techniques, lots of pictures, moving images, excessive capitalization, ad hominem attacks, emotionally charged words and pictures, unsafe generalizations and other logical fallacies.
  o   WikiLeaks. These tweets provide links to unverified claims or irrelevant documentation on WikiLeaks.org.
  o   Russia. This content was produced by known Russian sources of political news and information.

- Other Political News and Information
  o   Citizen, Civic, or Civil Society. Links to content produced by independent citizens, civic groups, or civil society organizations. Blogs and websites dedicated to citizen journalism, citizen-generated petitions, personal activism, and other forms of civic expression that display originality and creation more than curation or aggregation.
  o   Humor and Entertainment. Content that involves political jokes, sketch comedy, political art or lifestyle- or entertainment-focused coverage.
  o   Religion. Links to political news and information with distinctly religious themes and faith-based editorializing presented as political news or information.
  o   Other Political Content. Myriad other kinds of political content, including portals like AOL and Yahoo! that do not themselves have editorial policies or news content, survey providers, and political documentary movies.

- Other
  o   Social Media Platforms. Links that simply refer to other social media platforms, such as Facebook or Instagram. If the content at the ultimate destination could be attributed to another source, it is.
  o   Not Available. Links that are no longer available or not successfully archive after repeated attempts.
  o   Other Non-Political. Links to sites that actually have no political content.

Our two-stage coding process involved developing an initial, grounded coding scheme and running it as a kind of pilot study of a subsample of Michigan-based users (See Memo 2017.1).[11] We then revised our definitions and recoded the complete dataset according to the categories defined below.

This methodology has several limitations. Tweets about the US Presidential Election by individuals who did not use one of these hashtags would not have been captured. Tweets from people who used these hashtags but were tweeting about something else would be captured in this sample. We infer that volunteered location information is a proxy for the political constituency of a human user. While demographic factors may explain a user's decision to geotag tweets, researchers outside Twitter know less about the distribution of self-reported location information.[12] There are some complex techniques for inferring physical location from tweet content or metadata, but there is evidence that volunteered information from urban areas is more trustworthy and it is not clear that the machine learning techniques yield better data than self-reported data.[13,14] The coding of source types was derived from the dataset and is not intended to be a comprehensive list of all types of information providers. Some types of political news and information are easy to catalogue. The category of "other content" was used for links to completely unpolitical pages. The overall percentages of different information sources are intended as a metric for the overall information environment surrounding the 2016 Presidential Election and as a way to stimulate further research and conversation.

**FINDINGS AND ANALYSIS**

This sample allows us to draw some conclusions about the character and process of political conversation over Twitter during the election, particularly as it relates to US voters and the circulation of different kinds of news and political information in swing states.

In Data Memos 2016.1-2016.4 we analyze political conversation over Twitter during US Presidential candidate debates and the 10 days leading up to the election itself.[15–18] Data Memo 2017.1 evaluates the circulation of junk news by taking a close look at what kinds of content Twitter users in Michigan were sharing just before the election.[11] In that study, we found that (1) in Michigan, conversation about politics over Twitter mirrored the national trends in that Trump-related hashtags were used more than twice as often as Clinton-related hashtags. (2) Social media users in Michigan shared a lot of political content, but the amount of professionally researched political news and information was consistently smaller than the amount of extremist, sensationalist, conspiratorial, masked commentary, fake news and other forms of junk news. (3) Not only did such junk news "outperform" real news, but the proportion of professional news content being shared hit its lowest point the day before the election. If junk news was so prevalent in this important swing state, how was it distributed across the rest of the country?

*What Were Citizens Sharing?* To investigate the distribution of political news and information being shared over social media we took the catalogue of content found in Michigan and classified the links being shared in the entire national sample. Table 1 catalogues the different kinds of URLs being shared in election-related tweets by users in the US. We removed tweets with links that were in foreign languages, no longer available, or inaccessible for some other reason. All in all, 1,033,742 links were still available online and successfully labelled.

Table 1 presents the findings of this grounded catalogue of content. Overall, 20% of the links being shared with election-related hashtags came from professional news organizations. Links to content produced by government agencies, political parties and candidates, or experts, altogether added up to just 10% of the total. Indeed, only small fractions of the content being shared originated with the political parties, candidates, civil society groups, universities or public agencies.

Two things should be noted across categories. First, the number of links to professionally produced content is less than the number of links to polarizing and conspiratorial junk news. In other words, the number of links to Russian news stories, unverified or irrelevant links to WikiLeaks pages, or junk news was greater than the number of links to professional researched and published news. Indeed, the proportion of misinformation was twice that of the content from experts and the candidates themselves. Second, a worryingly large proportion of all the successfully catalogued content provides links to polarizing content from Russian, WikiLeaks, and junk news sources. This content uses divisive and inflammatory rhetoric, and presents faulty reasoning or misleading information to manipulate the reader's understanding of public issues and feed conspiracy theories. Thus, when links to Russian content and unverified WikiLeaks stories are added to the volume of junk news, fully 32% of all the successfully catalogued political content was polarizing, conspiracy driven, and of an untrustworthy provenance.

Table 1: What Political News and Information Were US Twitter Users Sharing during the Election?

| Type of Source | N | % of Subset | % of Total |
|---|---|---|---|
| **Professional News Content** | | | |
| Major News Brands | 196,697 | 77 | |
| Minor News Brands | 60,055 | 23 | |
| Subtotal | 256,752 | 100 | 20 |
| **Professional Political Content** | | | |
| Political Party or Candidate | 121,323 | 93 | |
| Experts | 6,350 | 5 | |
| Government | 2,294 | 2 | |
| Subtotal | 129,967 | 100 | 10 |
| **Polarizing and Conspiracy Content** | | | |
| Junk News | 203,591 | 79 | |
| WikiLeaks | 48,068 | 19 | |
| Russia | 7,683 | 3 | |
| Subtotal | 259,342 | 100 | 20 |
| **Other Political News and Information** | | | |
| Other Political | 97,900 | 59 | |
| Citizen, Civic or Civil Society | 34,935 | 21 | |
| Political Humor or Entertainment | 30,021 | 18 | |
| Religion | 2,124 | 1 | |
| Political Merchandise | 2,067 | 1 | |
| Subtotal | 167,047 | 100 | 13 |
| **Other** | | | |
| Social Media Platform | 195,470 | 42 | |
| Not Available | 241,688 | 52 | |
| Other Non-Political | 25,164 | 5 | |
| Subtotal | 462,322 | 100 | 36 |
| **TOTAL** | 1,269,736 | | 100 |

*Source: Authors'* calculations from data sampled 1-11/11/16. For the full list of hashtags used to capture this sample see Memo 2017.1).[11]

***Polarizing Content and Swing States***. To help understand the national distribution of polarizing content from Russian, WikiLeaks, and junk news sources, we then create a simple index, organized by states. Since states have different numbers of Twitter users, for each state we calculated the ratio of a state's tweets about politics to the entire country's tweets about politics. Then we added up all the tweets with junk news, links to unverified WikiLeaks pages, or links to Russian content (such as RussiaToday or Sputnik). For each state, we calculated a second ratio—the ratio of each state's junk content to the entire country's junk content. Putting these two ratios together gives an index of the proportion of junk news tweets from the state relative to the number of all political tweets from the state. The distribution of this "ratio of ratios" is uneven because half the distribution of possible values ranges from 0 to 1 (less junk news than expected given national averages) and the other half ranges from 1 to +infinity (more junk news than expected given national averages). However, by taking the natural log of the ratio of ratios the index will become more balanced: from –infinity to 0 becomes less junk news than expected, and 0 to +infinity becomes more junk news than expected. Mathematically, the 0 point on the index is the national average.

$$Ratio\ of\ Ratios_T = \frac{\frac{PolarizingLinks_{state}}{\sum_{all\ states} PolarizingLinks}}{\frac{PoliticalLinks_{state}}{\sum_{country} PoliticalLinks}}$$

Expression A: Indexing Polarizing Political Content from Russian, WikiLeaks, and Junk News Sources

Table 2 identifies the weighted junk news scores for each state. States with scores around zero are close the national average, so they have about the concentration of junk news expected given the amount of Twitter conversation about politics occurring in that state. States with more negative scores have less junk news than expected given the amount of Twitter conversation about politics occurring in that state. Finally, states with more positive scores have unusual concentrations of junk news, even considering how much political conversation was occurring among Twitter users in that state. States that were considered swing states on November 2, 2017, according to the non-partisan National Constitution Center are marked by an (*) asterisk.[19]

At the bottom of this table the simple but revealing averages are calculated. This index is weighted for the relative number of tweets about politics generated by each state. If a state had about the level of junk news to be expected for the size of its Twitter user base, the state earned a score around the national average. If it had more junk then expected it scored above that, and if it had less than expected it scored below that. Table 2 reveals that polarizing content was surprisingly concentrated in swing states even considering the amount of political conversation occurring in the state. Indeed, 12

Table 2: Concentration of Polarizing Political Content from Russian, WikiLeaks, and Junk News Sources, Indexed by State

| State | Junk News Index |
|---|---|
| North Dakota | -0.75 |
| Washington D.C. | -0.54 |
| Iowa* | -0.38 |
| New York | -0.30 |
| Wyoming | -0.26 |
| New Mexico | -0.20 |
| Nebraska | -0.19 |
| Rhode Island | -0.18 |
| Hawaii | -0.17 |
| Wisconsin* | -0.16 |
| Minnesota* | -0.15 |
| Washington | -0.14 |
| Illinois | -0.13 |
| Kansas | -0.13 |
| Connecticut | -0.12 |
| Utah | -0.11 |
| Idaho | -0.09 |
| Mississippi | -0.07 |
| Alaska | -0.07 |
| Massachusetts | -0.06 |
| California | -0.03 |
| Maine* | -0.03 |
| Oregon | -0.02 |
| Louisiana | -0.01 |
| Colorado* | +0.01 |
| Ohio* | +0.01 |
| New Jersey | +0.01 |
| Maryland | +0.03 |
| Michigan* | +0.05 |
| Arkansas | +0.05 |
| Texas | +0.06 |
| Georgia* | +0.07 |
| South Carolina | +0.10 |
| Vermont | +0.10 |
| New Hampshire* | +0.11 |
| Pennsylvania* | +0.12 |
| Oklahoma | +0.14 |
| North Carolina* | +0.14 |
| Virginia* | +0.14 |
| Florida* | +0.16 |
| Nevada* | +0.18 |
| Indiana | +0.18 |
| Tennessee | +0.20 |
| Missouri* | +0.22 |
| Alabama | +0.23 |
| Arizona* | +0.25 |
| Kentucky | +0.26 |
| South Dakota | +0.27 |
| Delaware | +0.35 |
| Montana | +0.45 |
| West Virginia | +0.47 |
| | |
| Averages | |
| Not Contested State Average | -0.02 |
| National Average | +0.00 |
| Swing State Average | +0.05 |

Source: Authors' calculations from data sampled 1-11/11/16.
Note: (*) Indicates a swing state in November 2016 according to the National Constitution Center.

of the 16 swing states were above the average among less competitive states, and 11 of the 16 were above the national average. States that were not hotly contested had, on average, lower levels of junk news.

Figure 1: States With Above Average Concentrations of Polarizing Political Content from Russian, WikiLeaks, and Junk News Sources

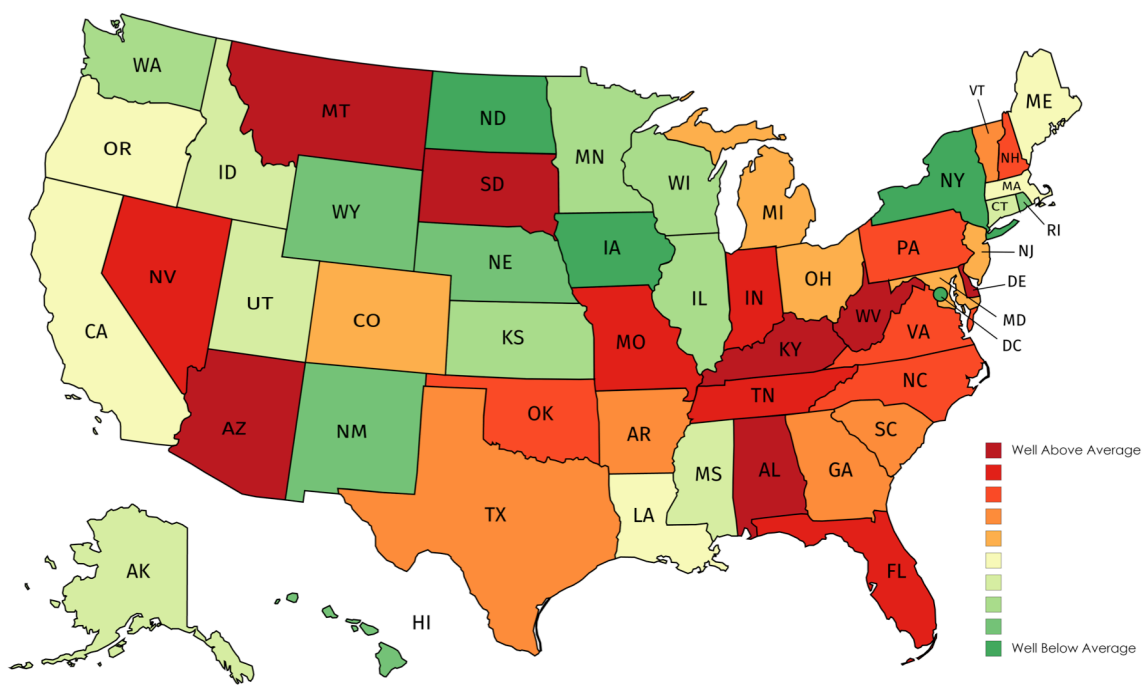

*Source: Authors' calculations from data sampled 1-11/11/16.*
*Note: Shades of green indicate that state users shared less of this content than expected given the size of their Twitter user base. Shades of red indicate that state users shared more junk content than expected given the size of their Twitter user base.*

## CONCLUSIONS

Many of the swing states getting highly concentrated doses of polarizing content were also among those with large numbers of votes in the Electoral College. Figure 1 presents a heat map of the distribution of junk news around the country.

The term "fake news" is difficult to operationalize, so our grounded typology reflects the diversity of organizations behind the content that was circulated over Twitter by people in the United States in the ten days before voting day. Social media users in many states traded links to high quality political news and information. Junk news, characterized by ideological extremism, misinformation and the intention to persuade readers to respect or hate a candidate or policy based on emotional appeals, was just as, if not more, prevalent than the amount of information produced by professional news organizations.

## ABOUT THE PROJECT

The Project on Computational Propaganda (www.politicalbots.org) involves international, and interdisciplinary, researchers in the investigation of the impact of automated scripts—computational propaganda—on public life. *Data Memos* are designed to present quick snapshots of analysis on current events in a short format. They reflect methodological experience and considered analysis, but have not been peer-reviewed. *Working Papers* present deeper analysis and extended arguments that have been collegially reviewed and that engage with public issues. The Project's articles, book chapters and books are significant manuscripts that have been through peer review and formally published.

## ACKNOWLEDGMENTS AND DISCLOSURES

The authors gratefully acknowledge the support of the European Research Council, "Computational Propaganda: Investigating the Impact of Algorithms and Bots on Political Discourse in Europe," Proposal 648311, 2015-2020, Philip N. Howard, Principal Investigator. Project activities were approved by the University of Oxford's Research Ethics Committee. The project gratefully thanks the Ford Foundation for their support. Any opinions, findings, and conclusions or recommendations expressed in this material are those of the authors and do not necessarily reflect the views of the University of Oxford or the European Research Council.